\documentstyle[11pt,newpasp,twoside,epsf]{article}
\def\edcomment#1{\iffalse\marginpar{\raggedright\sl#1\/}\else\relax\fi}
\reversemarginpar

\begin{document}
\title{The Origins of Disk Heating}
 \author{Michael R. Merrifield}
\affil{School of Physics \& Astronomy, University of Nottingham, 
       Nottingham, NG7 2RD, England}
\author{Joris Gerssen and Konrad Kuijken}
\affil{Kapteyn Institute, PO Box 800, 9700 AV Groningen, The Netherlands}

\begin{abstract}
By making spectral absorption-line observations of disk gal\-axies at
intermediate inclinations, we have been able to determine the
amplitude of their constituent stars' random motions in three
dimensions.  This full measure of the shape of the velocity ellipsoid
is a useful diagnostic for determining the ``heating'' mechanism
responsible for creating the random motions.  The analysis implies
that the main heating process varies with galaxy type, with minor mergers
dominating the heating in early-type disk galaxies, and spiral density
waves the most important mechanism in late-type systems.

\end{abstract}

\section{Introduction}

Observations of edge-on galaxies reveal that they are remarkably thin
structures, but not infinitely so.  Their finite extent perpendicular
to the disk plane can be attributed to the excursions that stars take
in this direction due to their random motions.  Stars are born in a
very thin layer of gas in these galaxies, which, due to its
collisional nature, has very little by way of random motions.  Thus,
stars must acquire their random velocities in later life, and several
candidates have been suggested for the source of this ``heating.''
One way that a star can acquire such a random component is by
scattering gravitationally from a more massive object, such as a giant
molecular cloud.  Similarly, the mass enhancements associated with
spiral density waves can gravitationally scatter stars.  Finally, the
heating could be due to an external source, such as a small satellite
galaxy colliding with the disk; the energy dissipated in such a
collision would significantly heat the original stellar population.

How, then, does one distinguish between these possibilities?  One very
useful diagnostic is provided by the relative amplitudes of the random
motions in different directions (Jenkins \& Binney 1990).  Scattering
off giant molecular clouds is a very stochastic process, so heats the
population reasonably isotropically.  Density waves, on the other
hand, are encountered in a more predictable fashion.  Since two-armed
spirals tend to be most common, a star will receive a kick from the
associated density wave twice per orbit.  This frequency is close to
the star's natural oscillation frequency in the radial direction, so
random motions in this direction will be increased rapidly by this
near-resonant process, making the radial velocity dispersion,
$\sigma_R$, significantly greater than the velocity dispersion
perpendicular to the plane, $\sigma_z$ (Jenkins \& Binney 1990).
Finally, mergers with satellites couple closely to oscillations that
can be excited perpendicular to the disk plane, so such heating events
would enhance $\sigma_z$ relative to $\sigma_R$ (Sellwood, Nelson \&
Tremaine 1998).  Thus, measuring the ratio of
$\sigma_z/\sigma_R$ offers a simple test for determining the cause of
the random motions.

The main complication in applying this diagnostic in practice is that
it is difficult to measure $\sigma_R$ and $\sigma_z$ simultaneously.
Since Doppler broadening of spectral lines only enables one to
determine the line-of-sight velocity dispersions of external galaxies,
$\sigma_z$ is usually only measured for face-on galaxies while
$\sigma_R$ is only observed in edge-on systems (e.g.~van der Kruit \&
Freeman 1986).  The only exception to this problem is the Milky Way,
where one can measure the full three-dimensional motions of nearby
stars from their Doppler shifts and proper motions.  

However, we have now shown that one can also determine the random
velocities in all three dimensions for some external galaxies.  If one
picks a disk galaxy at intermediate inclination, then observations of
the line-of-sight velocity dispersion along its minor axis are a
combination of $\sigma_R$ and $\sigma_z$, while measurements along its
major axis reveal a combination of $\sigma_z$ and the remaining
component, $\sigma_\phi$, the velocity dispersion in the tangential
direction.  One then invokes the fact that these quantities are also
related by the equations of stellar dynamics (Binney \& Tremaine
1987); specifically, the asymmetric drift equation relates $\sigma_R$
to the mean rotation speed of the stars, $\overline{v}$, and the local
circular speed, $v_c$.  The former of these quantities can be obtained
from the Doppler shift in spectral absorption lines along the major
axis, while the latter can be obtained from the Doppler shifts in any
emission lines, which arise from gas on circular orbits.  For an
exponential disk of scalelength $h_{\rm d}$, the asymmetric drift
equation can be written
\begin{equation}
v_c^2 - \overline{v}^2 = \sigma_R^2\left[{R \over h_{\rm d}}
                       - R {\partial \over \partial R} \ln(\sigma_R^2) 
                       - {1 \over 2} 
                       + {R \over 2v_c}{\partial v_c \over \partial R} 
                       \right].
\end{equation}
Combining this equation with the two observable quantities (the
line-of-sight kinematics along the major and minor axes), one can
solve for all three components of the velocity dispersion, $\sigma_R$,
$\sigma_\phi$, and $\sigma_z$.  

Even with the best data available today, one cannot perform this
analysis in a completely non-parametric fashion.  Instead, one adopts a
somewhat simplified model in which the ratios of the dispersions
remain fixed.  By solving for the free parameters in this model, one in
essence obtains average values for ratios such as
$\sigma_z/\sigma_R$, and hence a measure of what globally is the most
important heating mechanism in the galaxy.

\section{Results}

As mentioned above, by looking at nearby stars, one can determine all
three components of the velocity dispersion of the Milky Way in the
solar neighbourhood.  This analysis reveals a ratio of
$\sigma_z/\sigma_R \sim 0.53$ (Dehnen \& Binney 1998).  Such a low
value is consistent with what one would expect for a galaxy in which
spiral density waves are the major heating mechanism.

\begin{figure}
\plotfiddle{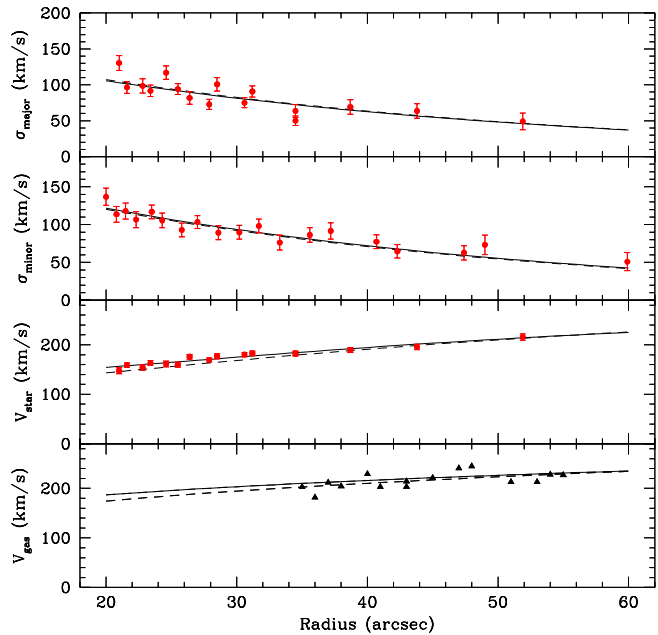}{3.1in}{0}{45}{45}{-135}{-75}
\caption{Plot showing the kinematic data obtained for NGC~488, and the
resulting model fits.  The panels show the line-of-sight velocity
dispersions along the major and minor axes, and the stellar and
gaseous rotation rates along the major axis.}
\end{figure}

By observing kinematics along the major and minor axes of NGC~488, and
using the technique described in the previous section, we made the
first comparable measurement for an external galaxy (Gerssen, Kuijken
\& Merrifield 1997).  Fitting to these data (shown in Fig.~1) reveals
a higher ratio of $\sigma_z/\sigma_R \sim 0.7$.  This galaxy is of an
earlier type than the Milky Way (Sb as opposed to Sbc), with
correspondingly weaker spiral structure.  It is also bright in CO
emission (Young et al.\ 1995), implying that it is rich in molecular
clouds.  It is therefore not too surprising that the
$\sigma_z/\sigma_R$ diagnostic favours molecular clouds over density
waves as the source of the heating in this case.

More recently, we have made a similar analysis of the Sab galaxy
NGC~2985 (Gerssen, Kuijken \& Merrifield 2000).  The even earlier type
of this galaxy implies that spiral density waves are unlikely to have
a significant role to play, but the absence of strong CO emission also
makes molecular clouds implausible sources of disk heating.  It is
therefore interesting that the analysis of this galaxy returns a ratio
of $\sigma_z/\sigma_R \sim 0.85$.  Such an extreme value strongly
favours the third possibility, mergers with satellites, as the heating
mechanism.

\begin{figure}
\plotfiddle{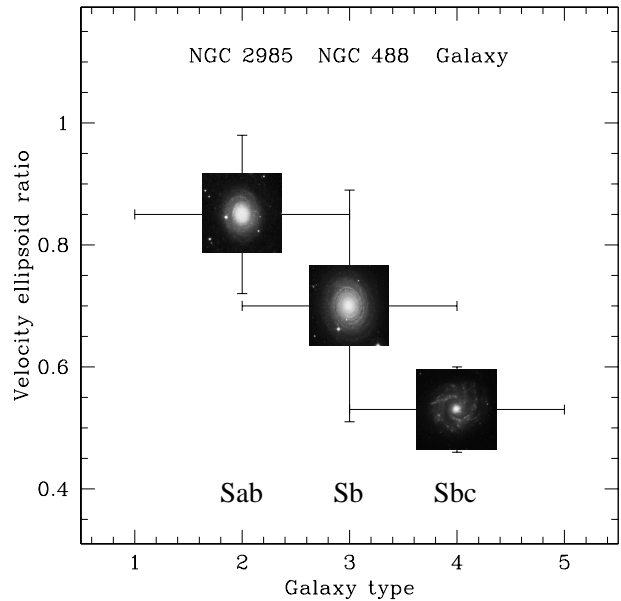}{2.95in}{0}{45}{45}{-195}{-20}
\caption{Plot showing the velocity dispersion ratio,
$\sigma_z/\sigma_R$, as a function of Hubble type for the three
galaxies for which this quantity has now been determined.  The error
bars show the uncertainties in these two quantities, and the ``postage
stamp'' images at the centres of each error bar illustrate the
galaxies' morphologies.}
\end{figure}

With three measurements, we can start to look for any systematic
trends.  As Fig.~2 shows, there is some evidence for such a trend: the
earlier the Hubble type of the galaxy, the larger the value of
$\sigma_z/\sigma_R$.  Galaxy interactions provide a plausible
explanation for this correlation: in addition to enhancing the value
of $\sigma_z/\sigma_R$, mergers are likely to transform galaxies to
earlier Hubble types.  However, the large error bars on Fig.~2 mean
that any such interpretation should be viewed with caution.  The
modelling process required to calculate $\sigma_z/\sigma_R$ is quite
complex and results in a rather uncertain estimate, and even the
definition of a galaxy's Hubble type has significant uncertainty due
to the subjectivity of the classification.  Thus, the the tight
correlation between these quantities must be somewhat fortuitous.
Clearly, $\sigma_z/\sigma_R$ needs to be measured for a larger sample of
galaxies before any general conclusions can be drawn, but even these
preliminary results show the promise of this technique for uncovering
the origins of disk heating.

\end{document}